\documentclass[twocolumn,showpacs,preprintnumbers,amsmath,amssymb,prb]{revtex4}
\usepackage{graphicx}
\usepackage{dcolumn}
\usepackage{bm}
\usepackage{epstopdf}

\usepackage{color}

\begin{document}

\title{
Interacting spinless fermions on the square lattice: Charge order, phase separation, and superconductivity }

\author{Kou-Han Ma}
\affiliation{Department of Physics, Renmin University of China, 100872 Beijing, China}
\author{Ning-Hua Tong}
\email{nhtong@ruc.edu.cn}
\affiliation{Department of Physics, Renmin University of China, 100872 Beijing, China}
\date{\today}

\begin{abstract}
We investigate the phase diagram of spinless fermions on a square lattice with nearest-neighbor interaction, using the recently developed projective truncation approximation in Green's function equation of motion. 
For attractive interaction, the ground state is in a homogeneous $p + ip$ superconducting (SC) phase at high or low fermion
 densities. Near half filling is a phase separation (PS) between the SC phases. Allowing inhomogeneous solution, we obtain $p$-wave SC domains with positive interface energy. As temperature increases, the SC phases transit into normal phases above $T_{sc}$, generating a homogeneous normal phase (far away from $n=1/2$), or a PS between normal phases with different densities (close to $n=1/2$). Further increasing temperature to $T_{ps}$, the PS disappears and the particle-hole symmetry of the Hamiltonian is recovered. 
For repulsive interaction, depending on the filling, the ground state is in charge-ordered phase (half filling), charge-disordered phase (large hole/particle doping), or PS between them (weak doping). At finite temperature, the regime of charge order phase moves to finite $V$ and extends to finite doping regime. 

\end{abstract}
\pacs{71.10.Fd, 64.70.Tg, 74.20.Rp, 74.81.-g}


\maketitle

\begin{section}{Introduction}

The spinless fermion (SF) model is a simple but important model in quantum many-body physics. Historically it originates from the study of metal-insulator transition.$\cite{WK15}$ In the early stage, this model was used to describe a class of materials with Verwey transition,$\cite{JC16,DI16.1,FW16.2}$ as well as the thermodynamic and transport properties of superionic conductors.$\cite{LL16.3,SG16.4}$ It can be used to describe the phase separation, stripe order, and nematic order in cuprates and organic superconductors.$\cite{HS16.5,RD16.6,CH32,SK16.7}$ In recent years, this model has wide applications in the emerging fields such many-body localization,$\cite{PS16.8,MF16.8.5,YL16.9}$ charge fractionalization,$\cite{FP17.1,AO17.2}$ time-reversal symmetry breaking,$\cite{OT20,SR17,SC18,SC19}$ quantum criticality,$\cite{LW21,ZL21,YT19.2}$ quantum quench,$\cite{BD19.1}$ matter-field interaction,$\cite{DGC19.1.1, UB19.1.2}$ and topological states.$\cite{LL19.3}$

One of the simplest forms of the SF model is defined on a bipartite lattice with only nearest-neighbor hopping and density-density interaction. The Hamiltonian reads
\begin{equation}\label{eq:1}
	H = -t\sum_{\langle ij \rangle}(c_i^{\dagger} c_j + h.c.)+V\sum_{ \langle ij \rangle}n_in_j - \mu \sum_i n_i.
\end{equation}
Here $c_i^{\dagger}$ ($c_i$) is the fermion  creation (annihilation) operator on site $i$. $n_i=c_i^{\dagger} c_i$ and $\langle ... \rangle$ represents the nearest-neighbor pair summation. On bipartite lattice, this model has particle-hole (PH) symmetry at $\mu=zV/2$, where $z$ is the coordination number. In this work, $t = 1.0$ is taken as the unit of energy and periodic boundary condition is used.

The properties of this system has been well studied for the cases of one spatial dimension and infinite spatial dimensions.
In one dimension, by Jordan-Wigner transformation, Eq.($\ref{eq:1}$) can be mapped into a spin-$1/2$ XXZ model under Zeeman field.$\cite{EF22}$ At half filling that corresponds to zero field of the spin model, the ground state is either a Luttinger liquid ($-2t < V \leq 2t$) or a phase with charge order (CO) ($V>2t$).$\cite{FH23,MT23,EJ24}$ In infinite dimensions, Uhrig and Vlaming$\cite{GU25}$ obtained the full phase diagram and observed an incommensurate phase.

For other spatial dimensions, this model has been studied using the analytical methods such as perturbation theory,$\cite{EH26}$ perturbative-variational approach,$\cite{AM27}$ Hartree-Fock approximation,$\cite{WC28,JW29}$ as well as numerical techniques including cluster approximation,$\cite{BL29,DI29}$ quantum Monte Carlo,$\cite{DS30,JG31}$ exact diagonalization,$\cite{CH32}$ fermionic projected entangled-pair states,$\cite{PC33}$ and variational Monte Carlo.$\cite{JS34,OS35,JS36}$. One of the basic issues is the phases and their stability.
For half filling and $V>0$, the stability of the CO ground state is discussed$\cite{PL36.1}$ and the ground state degeneracy~$\cite{ZW36.2}$ analysed  by strict proof. The van Hove singularity at Fermi energy facilitates both PS and superconductivity (SC) for $V<0$, as analysed by random phase approximation$\cite{JG31}$ and Bogoliubov mean-field approximation.$\cite{MC37}$ Various ordering is often accompanied with phase separation (PS), i.e., the tendency of particles to segregate into inhomogeneous state in real space.

Surprisingly, despite extensive studies in the past decades, a complete phase diagram containing charge order ($V>0$ case), superconductivity ($V<0$ case), and PS is still absent for this model. The interplay between PS and various orderings, e.g., the CO and SC order studied in this paper, is still awaiting a deeper understanding, especially for small $V$ regime. 
Ordering and PS are prevalent phenomena in many strongly correlated electron materials ranging from colossal magneto-resistance manganite\cite{ED1}, to organic superconductors\cite{TS1}, and to high temperature superconducting cuprates.\cite{TM1} The general understanding of this issue acquired by studying the simple SF model could benefit the study of other more complicated systems.

In this paper, we address this problem using the Green's function (GF) equation of motion (EOM) method with projective truncation approximation (PTA). The robustness of our conclusion is examined by expanding the operator basis beyond the mean-field level. We map out the global phase diagram which contains CO, SC, and PS phases. The interplay between phase separation and the ordering of spinless fermions (i.e., SC and CO) is elucidated.   

Our findings are the following. For repulsive interaction, depending on the filling, the ground state of Eq.(1) is in CO phase (half filling), PS between CO and the charge-disordered (CD) phase (weak doping), or CD phase (large doping). At finite temperature, the regime of CO phase moves to finite $V$ and extends to finite doping regime. For attractive interaction, at $T=0$, any weak attraction produces fermion pairing and leads to $p_x$ + $ip_y$ SC phase. Depending on the filling of spinless fermions, the ground state is either in a homogeneous SC phase (far away from half filling) or PS between them (near half filling). The inhomogeneous solution gives $p$-wave SC domains in real space with positive interface energy. 

With increasing temperature, the SC states first transit into normal phases at a lower temperature $T_{sc}$, and then PS disappears at a higher $T_{ps}$, showing successive recovery of $U(1)$ and PH symmetry. In both cases of positive and negative $V$, we find that PS occupies a significant portion of the phase diagram around half filling. It competes with CO ($V>0$) and SC($V<0$) on the thermodynamical level, i.e., PS suppresses the regions of two ordered states by tuning the density of spinless fermion away from the favourable level for ordering.

This paper is arranged as follows. For the sake of completeness, in Sec.II, we briefly introduce PTA in GF EOM. In Sec.III, the formalism of PTA for SF model using different operator bases is presented. In Sec.IV, we summarize the formula of Hartree-Fock-Bogoliubov (HFB) mean-field theory which is equivalent to PTA under the simple basis. Section.V presents our numerical results and analysis. Summary and discussion are given in Sec.VI.

\end{section}

\begin{section}{Introduction to GF EOM PTA}

In this section, we briefly introduce the method that we use to study SF model in this work, i.e., the GF EOM PTA method. The two-time GF EOM$\cite{PJ1,NS2,ST3,DZ4}$ is a traditional tool for studying quantum many-body problems. Its modern application, however, is hampered by the arbitrariness and uncontrolled nature of the truncation approximation.$\cite{ST3,JH5,CL6}$
Recently, based on the ideas of operator projection,$\cite{HM7,RZ8,RZ9,RZ10,YT11,LR12,DR12,ML13}$ Fan {\it et al}.$\cite{PF14}$ developed the systematic truncation scheme known as PTA to solve the GF EOM. With this method, GF with correct analytical structure can be obtained with controlled precision for a general quantum many-body system.$\cite{PF14p}$

For a given Hamiltonian $H$, we select a set of linear independent operators to form the vector $\vec{A}=( A_1, A_2,. .., A_n )^T$, which is supposed to include the most relevant excitations of the problem. A matrix of two-time retarded Fermion-type GF is defined as 
\begin{equation}      \label{eq:2}
 {\bf G} \left( \vec{A}(t) \Big| \vec{A}^{\dag} (t^{\prime}) \right) = -\frac{i}{\hbar} \theta(t-t^{\prime}) \left\langle \left\{ \vec{A}(t),\vec{A}^{\dag}(t^{\prime})  \right\}\right\rangle.
\end{equation}
Here, $\theta(t-t^{\prime})$ is the Heaviside step function. $\vec{A}(t)$ is the vector of basis operators in Heisenberg picture. The curly bracket represents anti-commutator. $\langle \hat{O} \rangle=Tr(e^{-\beta H}\hat{O})/ Tr e^{-\beta H}$ is the thermodynamical average of operator $\hat{O}$. Below, we take the natural unit, $\hbar = 1$.

In the frequency domain, the GF matrix satisfies the EOM
\begin{eqnarray}      \label{eq:3&4}
\omega G\left( \vec{A} \Big| \vec{A}^{\dag} \right)_{\omega} &=& \langle \{ \vec{A}, \vec{A}^{\dag} \} \rangle + G\left( [ \vec{A}, H ] \Big|\vec{A}^{\dag} \right)_{\omega} ,   \\
\omega G\left( \vec{A}\Big|\vec{A}^{\dag} \right)_{\omega} &=& \langle \{ \vec{A},\vec{A}^{\dag} \} \rangle - G\left( \vec{A}\Big|[\vec{A}^{\dag}, H ] \right)_{\omega}.   
\end{eqnarray}
Here, the square bracket represents commutator. For an incomplete basis, the commutator $[\vec{A},\hat{H}]$ is not closed but generates new linearly independent operators. The EOM therefore involve higher order GFs. The idea of PTA is to project $[\vec{A},\hat{H}]$ to $\vec{A}$. We denote the commutator as
\begin{equation}      \label{eq:5}
[ A_{i}, H ] = \sum_{j} { \bf M}_{ji} A_{j} + B_{i},
\end{equation}
where $B_i \not \in \{ A_i \}$. ${\bf M}$ is called a naturally closed matrix.
PTA amounts to approximate $B_i$ as a linear combination $B_i \approx \sum_j N_{ji} A_j $ and determine $\bf N$ by projecting the equation to basis $\{A_i\}$. For this purpose, we choose the inner product
\begin{equation}      \label{eq:6}
   (A|B) \equiv \langle \{ A^{\dagger}, B \} \rangle
\end{equation}
that satisfies the requirements of linearity and positivity.
After projection, Eq.($\ref{eq:5}$) is approximated as 
\begin{equation}      \label{eq:7}
 [ \vec{A}, H ] \approx {\bf M}_{t}^{T} \vec{A},
\end{equation}
where $ {\bf M}_t \equiv \bf M + \bf N = {\bf I}^{-1} {\bf L}$.
Here, the inner product matrix $\bf I$ is defined as ${\bf I}_{ij} \equiv  (A_{i} | A_{j})$. The Liouville matrix $\bf L$ is given by ${\bf L}_{ij} \equiv (A_{i}| [A_{j}, H])$. Both $\bf I$ and $\bf L$ are Hermitian and $\bf I$ is positive definite. This property ensures that the obtained approximate GF has only real simple poles.

Combining Eqs. ($\ref{eq:3&4}$) and ($\ref{eq:7}$) , we get the approximate GF matrix 
\begin{equation}      \label{eq:8}
G( \vec{A}|\vec{A}^{\dagger} )_{\omega} \approx \left( \omega {\bf 1} - {\bf M}_{t}^{T} \right)^{-1} { \bf I}^{T}.
\end{equation}
An equivalent expression reads
\begin{equation}      \label{eq:9}
G( \vec{A}|\vec{A}^{\dagger} )_{\omega} \approx ( {\bf IU} )^{\ast} \left( \omega {\bf 1} - {\Lambda}\right)^{-1} ({\bf IU})^{T},
\end{equation}
with $\bf U$ being the eigenvector matrix of the generalized eigen value problem ${\bf L U} = {\bf I U} \Lambda$. It fulfils the generalized orthogonal relation ${\bf U^{\dagger}IU=1}$. $\Lambda = \text{diag}(\lambda_1, \lambda_2, ..., \lambda_n)$ is a real diagonal matrix. This formal solution of $\bf G_{\omega}$ involves $\bf I$ and $\bf L$ which contain unknown static averages. Those averages of the form $\langle A_{j}^{\dagger} A_{i} \rangle$ can be calculated self-consistently from GF by the spectral theorem,
\begin{equation}      \label{eq:10}
\langle A_{j}^{\dagger} A_{i} \rangle  =  \sum_{k} \frac{ ({\bf IU})^{\ast}_{ik} ({\bf IU})_{jk} }{e^{\beta\lambda_{k}} + 1},
\end{equation} 
or equivalently,
\begin{equation}      \label{eq:11}
\langle \vec{A^{\dagger}} \vec{A}^T \rangle  =  \mathbf{I}(e^{\beta \mathbf{M}_t} + {\bf 1})^{-1}.
\end{equation}

For those averages that cannot be expressed in the form $\langle A_{j}^{\dagger} A_{i} \rangle$, additional approximation is required. If such averages appear in $\bf L$, Ref.\onlinecite{PF14} proposed the partial projection approximation (PPA). We first divide the basis into two subspaces $\{ A_i \}=\{ A^{(1)}_{i} \} \cup \{A^{(2)}_i\}$. Basis operator $A_{i}$ belongs to subspace $\{ A^{(1)}_i \}$ if $B_i =0$, and to subspace $\{ A^{(2)}_i \}$ if $B_i \neq 0$.  Accordingly, the matrix $\bf L$ and $\bf I$ become 2 by 2 block matrices. The idea of PPA is to approximate the projection of $B_i \neq 0$ to the subspace $\{ A^{(2)}_i \}$ by first projecting $B_i$ to $\{ A^{(1)}_i \}$ and then to $\{ A^{(2)}_i \}$, approximately expressing ${\bf L}_{22}$ in terms of $\bf I$ and $\bf M$. 
Employing the Hermiticity of ${\bf L}$, one gets
\begin{equation}       \label{eq:12}
\mathbf{L} \approx  \left(
\begin{array} {cc}
    \mathbf{ (IM) }_{11}  \,\,  & [ \mathbf{ (IM) }_{21}]^{\dagger}  \\
    \mathbf{ (IM) }_{21}  \,\,  & \frac{1}{2}[ \mathbf{ L}^{a}_{22} + (\mathbf{ L}^{a}_{22})^{\dagger} ]
\end{array} \right),
\end{equation}
where
\begin{equation}      \label{eq:13}
\mathbf{ L}^{a}_{22} = ({\bf IM})_{22} + {\bf I}_{21} [ {\bf I}_{11} ]^{-1} {\bf P}_{12},
\end{equation}
\begin{equation}      \label{eq:14}
 \mathbf{P}_{12} = \left[ \mathbf{ (IM) }_{21} \right]^{\dagger} - \mathbf{ (IM) }_{12}.
\end{equation}
If the inner product matrix $\bf I$ contains the averages outside the form $\langle A_{j}^{\dagger} A_{i} \rangle$, GF of the type $G(\vec{A}|O^{\dagger})_{\omega}$ needs to be calculated for properly chosen $\hat{O}$. Besides, PPA may break the PH symmetry of Hamiltonian. If that happens, we need to replace Eq.($\ref{eq:12}$) with the PH symmetry-conserving formalism~$\cite{PF14}$.
The GF obtained from the above procedure are guaranteed to obey the causality and energy conservation. It was confirmed on the Anderson impurity model that the precision of result improves systematically with enlarging basis size.$\cite{PF14p}$ 

\end{section}

\begin{section}{Application to spinless fermion model: formalism}

In this section, we apply EOM PTA to SF model Eq.($\ref{eq:1}$). 
The key is to select operator basis that contain 
the most relevant excitation operators of the system.
In this work, we consider the following bases. The simplest one is the $N$-dimensional basis of single-particle annihilation operators $\{ c_1, c_2, ..., c_N\}$. Here $N$ is the number of lattice sites. PTA with this basis is equivalent to Hartree-Fock (HF) mean-field approximation. We therefore call it HF basis. The second basis considered is $\{ c_i, \, n_{i+\delta}c_i\}$ ($i=1,2,...,N$), which includes the operators appearing from the commutator $[c_i, H]$. Here, $\delta$ stands for nearest-neighbor index, This basis has a dimension of $5N$ and is named p-5 basis (short for projection-5N dimension). These two bases cannot describe the superconducting phase. To take into account the superconducting order parameter, we extend the above bases by adding the Hermitian conjugate operators, forming the HFB basis $\{ c_i, \, c_i^{\dagger}\}$ ($i=1,2,...,N$) and the p-5-sc basis $\{c_i, \, c_i^{\dagger}, n_{i+\delta}c_i, \, n_{i+\delta}c_i^{\dagger}\}$ ($i=1,2,...,N$), respectively.

When the superconducting order parameter is zero (i.e., no U(1) symmetry breaking), the results from HFB and p-5-sc bases coincide with those from HF and p-5 bases, respectively. Besides the above four basis sets, in the case of superconducting state, we also consider a subspace of the p-5-sc basis, namely $\{c_i, \, c_i^{\dagger}, \, \sum_{\delta}n_{i+\delta}c_i, \, \sum_{\delta}n_{i+\delta}c_i^{\dagger} \}$ ($i=1,2,...,N$), which contains the operators in the commutator $[c_i, H]$ summed up as a single operator. We call this basis p-2-sc basis.

For the translation invariant phase, the calculation can be greatly simplified by using the translational symmetry of the Hamiltonian. To do that, we use the bases composed of the Fourier transform of the operators for each specific wave vector $\vec{k}$. It is worth noting that to meet the PH symmetry requirement, for the p-5, p-2-sc, and p-5-sc bases for a specific momentum $\vec{k}$, we put the two operators at a pair of momentums $\vec{k}$ and $(\pi, \pi)-\vec{k}$ into these bases. The reason is as follows.
For SF model Eq.($\ref{eq:1}$) on a square lattice, the PH transformation is defined as
\begin{equation}       \label{eq:34.1}
c_i^{\prime}=(-1)^i c_i^{\dagger}.
\end{equation}
At the PH symmetry parameter $\mu=2V$, the Hamiltonian is invariant under the PH transformation, i.e., $H^{\prime} = H$. 
To facilitate the consideration of PH symmetry, we define a  composite transformation as $\tilde{O} = (O^{\prime})^{\dagger}$ and require that the basis is invariant under this transformation. This in turn requires that both operators $O_{\vec{k}}$ and $O_{(\pi, \pi)-\vec{k}}$ be contained in the basis. Following the idea of Fan et al.,$\cite{PF14}$ we can then construct the PH symmetric natural closed matrix ${\bf M}$ and obtain the approximate GF matrix that satisfies PH symmetry. The details of constructing the PH symmetric natural closed matrix $\bf M$ is given in Appendix A.

In the following, for each operator basis used in this work, we give the corresponding matrices $\bf{I}$ and $\bf{M}$  (or the Liouville matrix $\bf{L}$ ).

\begin{subsection}{HF basis }
\begin{eqnarray} \label{eq:15}
\vec{A} = \left\{ \begin{array}{ll}
(c_1, \, c_2, \, ..., c_N )^{T},  & \textrm{real space,}\\
c_{\vec{k}}, & \textrm{momentum space.}
\end{array} \right.
\end{eqnarray}
Here, $c_{\vec{k}}=1/\sqrt{N} \sum_i e^{i\vec{k}\cdot\vec{r}_i}c_i$.
For this basis, PTA is equivalent to HF mean-field approximation. 
For the real space basis, matrices $\bf{I}$ and $\bf{L}$ are obtained as 
\begin{equation}      \label{eq:16}
\hspace{-1em}{\bf I}_{ij}=\delta_{ij}
\end{equation}
and
\begin{eqnarray}      \label{eq:17}
&&{\bf L}_{ij}= \nonumber \\
&& -t\sum_{\delta}\delta_{i,j+\delta} + V \sum_{\delta} \left(
\langle n_{i+\delta}\rangle \delta_{i,j}-\langle c_i^{\dagger}c_j\rangle \delta_{i,j+\delta} \right) -\mu \delta_{ij}, \nonumber \\
&&
\end{eqnarray}
respectively. In the momentum space basis, matrices ${\bf I}=1$ and $\bf{L}$ reads
\begin{equation}      \label{eq:19}
 L=\epsilon_{\vec{k}} -\mu + 4nV- \frac{V}{N}\sum_{\vec{k}', \delta} \cos\left[ (\vec{k}'-\vec{k})\cdot\vec{\delta} \right]
\langle c_{\vec{k}'}^{\dagger}c_{\vec{k}'}\rangle.
\end{equation}
Here, $\epsilon_k = -2t(\cos k_x+ \cos k_y)$ and $n=1/N \sum_
{\vec{k}}\langle c_{\vec{k}}^{\dagger}c_{\vec{k}}\rangle$ is the fermion density.

\end{subsection}

\begin{subsection}{p-5 basis}
The p-5 basis is defined as
\begin{eqnarray}   \label{eq:20}
&&  \vec{A} = \nonumber \\
&& \left\{ \begin{array}{ll}
(c_1, \, n_{1\delta_1 }c_1, \, n_{1\delta_2 }c_1, \, n_{1\delta_3 }c_1, \, n_{1\delta_4 }c_1, ... )^{T}, & \textrm{real space}\\
\left[c_{\vec{k}}, \, c_{\vec{\pi}-\vec{k}}, \, d_{\vec{k}}(\delta), \, d_{\vec{\pi}-\vec{k}}(\delta) \right]^{T}. & \textrm{momentum space} \\
\end{array} \right.   \nonumber \\
&&
\end{eqnarray}
Here, $n_{i \delta_z } = n_{i+\delta_z}$ ($z=1,2,3,4$), $d_{\vec{k}}(\delta)=1/\sqrt{N} \sum_i e^{i\vec{k}\cdot\vec{r}_i}n_{i+\delta}c_i$, and $\vec{\pi}=(\pi,\pi)$. 
To ensure the PH symmetry, we put operators with momentums $\vec{k}$ and $\vec{\pi}-\vec{k}$ into the basis together.
$\bf{L}$ contains the averages such as $\langle n_{i+\delta}c_{i+\delta'+\delta''}^{\dagger}c_{i+\delta'}\rangle$ and $\langle n_{i+\delta}n_{i+\delta'}n_{i+\delta''}\rangle$ which cannot be expressed as $\langle A_i^{\dagger} A_j\rangle$. We therefore use PPA to simplify the calculation.

For the real space basis, the matrix elements of $\bf{I}$ is given by the following equations,
\begin{align}  \label{eq:21}
& (c_i\mid c_j) = \delta_{ij}, \nonumber \\
& (c_i\mid n_{j+\delta}c_j) = \langle n_{i+\delta}\rangle
 \delta_{ij}-\langle c_{i}^{\dagger} c_{i-\delta}\rangle 
\delta_{j,i-\delta}, \nonumber \\
& (n_{i+\delta}c_i\mid n_{j+\delta'}c_j)  \nonumber \\
& = \langle n_{i+\delta}n_{i+\delta'}\rangle \delta_{ij} - \langle n_{i+\delta+\delta'}c_{i}^{\dagger} c_{i+\delta}\rangle \delta_{j,i+\delta} -\langle n_{i+\delta}c_{i}^{\dagger} c_{i-\delta'}\rangle \delta_{j,i-\delta'}. \nonumber \\
&&
\end{align}
The elements of the matrix $\bf M$ can be extracted from the following commutators,
\begin{align}  \label{eq:22}
		& [c_i, H]=-t\sum_{\delta}c_{i+\delta}
+V\sum_{\delta}n_{i+\delta}c_i - \mu c_i , \nonumber\\
		& [n_{i+\delta}c_i, H]= \nonumber \\
&		 -t n_{i}c_{i+\delta}+(V-\mu) n_{i+\delta}c_i - \frac{t}{2}\sum_{\delta'\not =\delta}c_{i+\delta'} \nonumber\\
		& + \frac{V}{2}\sum_{\delta'\not =\delta} \left(n_{i+\delta}c_i+n_{i+\delta'}c_i \right)+B_i(\delta).
\end{align}
Here,
\begin{align}      \label{eq:23}
\hspace{-8em}& B_i(\delta)= \frac{t}{2}\sum_{\delta'\not =\delta}c_{i+\delta'}-t\sum_{\delta'\not =\delta}n_{i+\delta}c_{i+\delta'}+\nonumber\\
&\hspace{3.5em} t\sum_{\delta'\not =-\delta}(c_{i+\delta+\delta'}^{\dagger}c_{i+\delta}-c_{i+\delta}^{\dagger}c_{i+\delta+\delta'})c_i \nonumber\\
& \hspace{1em} + V\sum_{\delta'\not =\delta}n_{i+\delta}n_{i+\delta'}c_i-\frac{V}{2}\sum_{\delta'\not =\delta}(n_{i+\delta}+n_{i+\delta'})c_i.
\end{align}
For the operator basis in momentum space, the inner product matrix $\bf{I}$ is given by
\begin{align} \label{eq:24}
		& (c_{\vec{k}}\mid c_{\vec{k}'})=\delta_{\vec{k},\vec{k}'}, \nonumber \\
		& (c_{\vec{k}}\mid d_{\vec{k}'}(\delta))= [n - \alpha_{\vec{k}}(\delta)]\delta_{\vec{k},\vec{k}'},\nonumber \\
&(d_{\vec{k}}(\delta)\mid d_{\vec{k}'}(\delta'))= \nonumber \\
& \left[ \frac{1}{N}\sum_i \langle n_i n_{i+\delta-\delta'}\rangle-\beta_{\vec{k}}(\delta,\delta')-\beta_{\vec{k}}^{\ast}(\delta',\delta)-\delta_{\delta,-\delta'}\alpha_{\vec{k}}^{\ast}(\delta) \right] \delta_{\vec{k},\vec{k}'}. \nonumber \\
&
\end{align}
The natural closure matrix $\bf{M}$ can be extracted from the following commutator relations:
\begin{align}  \label{eq:25}
   &  [c_{\vec{k}}, H]=(\epsilon_{\vec{k}}-\mu) c_{\vec{k}}+ V \sum_{\delta}d_{\vec{k}}(\delta),  \nonumber\\
   &  [d_{\vec{k}}(\delta), H] = \nonumber \\
   &-\frac{1}{2} \left(\epsilon_{\vec{k}}-t e^{-i\vec{k}\cdot \vec{\delta}} \right) c_{\vec{k}} + \left(\frac{5}{2}V-\mu \right) d_{\vec{k}}(\delta) \nonumber\\
& -t e^{-i\vec{k}\cdot \vec{\delta}} d_{\vec{k}}(-\delta) + \frac{V}{2} \sum_{\delta'\neq \delta}d_{\vec{k}}(\delta') + B_{\vec{k}}(\delta).
\end{align}
In the above equations,
\begin{align} \label{eq:26}
		& \alpha_{\vec{k}}(\delta)=\frac{1}{N}\sum_{\vec{k}'} e^{i(\vec{k}'-\vec{k})\cdot\vec{\delta}}\langle c_{\vec{k}'}^{\dagger} c_{\vec{k}'}\rangle, \nonumber \\
		&  \beta_{\vec{k}}(\delta,\delta')=\frac{1}{N}\sum_{\vec{k}'} e^{i(\vec{k}'-\vec{k})\cdot\vec{\delta}'}\langle d_{\vec{k}'}^{\dagger}(\delta) c_{\vec{k}'}\rangle,
\end{align}
and $B_{\vec{k}}(\delta)= 1/\sqrt{N} \sum_i e^{i\vec{k}\cdot\vec{r}_i}B_i(\delta)$.
For $\delta\neq\delta'$, the averages $\langle n_{i+\delta}n_{i+\delta'}\rangle$ in Eq.($\ref{eq:21}$) and
($\ref{eq:24}$) cannot be calculated self-consistently from the spectral theorem. We therefore used an additional decoupling approximation for it, $\langle n_{i+\delta}n_{i+\delta'}\rangle \approx \langle n_{i+\delta}\rangle \langle n_{i+\delta'}\rangle - 
\langle c_{i+\delta}^+c_{i+\delta'}\rangle \langle c_{i+\delta'}^+c_{i+\delta}\rangle$.
\end{subsection}

\begin{subsection}{HFB basis}

The simplest basis that is able to describe superconductivity is
\begin{displaymath} \label{eq:26.1}
\vec{A} = \left\{ \begin{array}{ll}
(c_1, c_2, ..., c_N, c_1^{\dagger}, c_2^{\dagger}, ..., c_N^{\dagger})^{T}, & \textrm{real space}\\
(c_{\vec{k}}, c_{-\vec{k}}^{\dagger})^{T}. & \textrm{momentum space}
\end{array} \right.
\end{displaymath}
For this basis, PTA is equivalent to HFB approximation. Analytical equations can be obtained and analysed for the momentum space basis, which will be left for the next section. For real space basis, we obtain the inner product matrix $\bf I = 1$. The matrix element of $\bf L$ reads
\begin{align}      \label{eq:26.2}
& \left( c_i \vert [c_j,H] \right)=-(c_j^{\dagger}|[c_i^{\dagger},H])  \nonumber \\
&= \hspace{0.5em}-t\sum_{\delta}\delta_{i,j+\delta}-\mu \delta_{ij}+V\sum_{\delta}\left( \langle n_{i+\delta}\rangle \delta_{i,j}-\langle c_i^{\dagger}c_j\rangle \delta_{i,j+\delta} \right) ,\nonumber \\
& \left( c_i^{\dagger}\vert [c_j,H]\right)=-\left(c_i \vert [c_j^{\dagger},H]\right)^{\ast} = V\sum_{\delta}\langle c_i c_j\rangle \delta_{i,j+\delta}.
\end{align}

\end{subsection}

\begin{subsection}{p-2-sc basis}
To take into account correlation effect, we enlarge HFB basis into the following $8$ dimensional operator basis
\begin{equation}      \label{eq.27}
  \vec{A}= \left( c_{\vec{k}}, \, c_{\vec{k}-\vec{\pi}}, \, c_{-\vec{k}}^{\dagger},\,  c_{\vec{\pi}-\vec{k}}^{\dagger}, \,  d_{\vec{k}},d_{\vec{k}-\vec{\pi}}, \, d_{-\vec{k}}^{\dagger}, \, d_{\vec{\pi}-\vec{k}}^{\dagger} \right) ^T.
\end{equation}
Here, $d_{\vec{k}}=\sum_{\delta}d_{\vec{k}}(\delta)$. Note that we have put the operators at momentums $\vec{k}$ and $\vec{k}-\vec{\pi}$ into the basis for PH symmetry reasons. Since the Liouville matrix $\bf{L}$ is complicated for this basis, we use PPA for it. Assigning the first four operators as block one, and the latter four as block two, we obtain the block matrices as
 $\mathbf{I}_{11}= \text{diag}(1,1,1,1)$, $\mathbf{I}_{21}=(\mathbf{I}_{12})^{\dagger}$, 
\begin{equation}       \label{eq:28}
\mathbf{I}_{12} = \left(
\begin{array} {cccc}
    4n-a_1(\vec{k})  \,\,  & 0  \,\,  & b_1(\vec{k}) \,\,  & 0  \\
     0  \,\,  & 4n+a_1(\vec{k})  \,\,  & 0 \,\,  & -b_1(\vec{k})  \\
    b_2(\vec{k})  \,\,  & 0  \,\,  & 4n-a_2(\vec{k}) \,\,  & 0  \\
    0  \,\,  & -b_2(\vec{k})  \,\,  & 0 \,\,  & 4n+a_2(\vec{k})  
\end{array} \right),
\end{equation}
and 
\begin{equation}       \label{eq:29}
\mathbf{I}_{22} = \left(
\begin{array} {cccc}
    -f_1(\vec{k})+h   & 0   & -g(\vec{k})   & 0  \\
     0    & f_1(\vec{k})+h    & 0   & g(\vec{k})  \\
    -g^{\ast}(\vec{k})    & 0    & -f_2(\vec{k})+h   & 0  \\
    0    & g^{\ast}(\vec{k})    & 0   & f_2(\vec{k})+h  
\end{array} \right).
\end{equation}
In the above two equations, the coefficients $a_{i}(\vec{k})$ and $b_{i}(\vec{k})$ ($i=1, 2$) are 
\begin{align} \label{eq:30}
		& a_1(\vec{k})=\frac{2}{N}\sum_{\vec{k}'} \left[\cos(k_x'-k_x)+\cos(k_y'-k_y)\right]\langle n_{\vec{k}'}\rangle,\nonumber \\
		& a_2(\vec{k})=\frac{2}{N}\sum_{\vec{k}'} \left[\cos(k_x'+k_x)+\cos(k_y'+k_y)\right]\langle n_{\vec{k}'}\rangle,\nonumber \\
		& b_1(\vec{k})=\frac{2}{N}\sum_{\vec{k}'} \left[\cos(k_x'+k_x)+\cos(k_y'+k_y)\right]\langle c_{\vec{k}'}^{\dagger} c_{-\vec{k}'}^{\dagger}\rangle,\nonumber \\
		& b_2(\vec{k})=\frac{2}{N}\sum_{\vec{k}'} \left[\cos(k_x'-k_x)+\cos(k_y'-k_y) \right]\langle c_{-\vec{k}'} c_{\vec{k}'}\rangle.
\end{align}
The coefficients $f_{1}(\vec{k})$, $f_{2}(\vec{k})$, and $g(\vec{k})$ are given as
\begin{align} \label{eq:30.1}
		& f_1(\vec{k})=\frac{2}{N}\sum_{\vec{k}'} \left[\cos(k_x'-k_x)+\cos(k_y'-k_y) \right] \nonumber \\
		&\hspace{6em} \left( \langle n_{\vec{k}'}\rangle+
		\langle d_{\vec{k}'}^{\dagger}c_{\vec{k}'}\rangle+
		\langle c_{\vec{k}'}^{\dagger}d_{\vec{k}'}\rangle \right),\nonumber \\
		& f_2(\vec{k})=\frac{2}{N}\sum_{\vec{k}'} \left[ \cos(k_x'+k_x)+ \cos(k_y'+k_y) \right]  \nonumber \\
		&\hspace{6em} \left( \langle n_{\vec{k}'}\rangle+
		\langle d_{\vec{k}'}^{\dagger}c_{\vec{k}'}\rangle+
		\langle c_{\vec{k}'}^{\dagger}d_{\vec{k}'}\rangle \right),\nonumber \\
		& g(\vec{k})=\frac{2}{N}\sum_{\vec{k}'} \left[ \cos k_x \cos k_x'+\cos k_y \cos k_y' \right] \langle d_{\vec{k}'}^{\dagger} c_{-\vec{k}'}^{\dagger}\rangle.\end{align}
In the diagonals of ${\bf I}_{22}$, the symbol $h$ is
\begin{equation}
 h=4n+\frac{1}{N}\sum_i\sum_{\delta}\sum_{\delta'\neq \delta}\langle n_i n_{i+\delta-\delta'}\rangle.
\end{equation}
It cannot be obtained directly by the spectral theorem. We therefore make an additional approximation $\langle n_{i}n_{i+\delta-\delta'}\rangle \approx \langle n_{i}\rangle \langle n_{i+\delta-\delta'}\rangle - 
\langle c_{i+\delta-\delta'}^{\dagger}c_{i}\rangle \langle c_{i}^{\dagger}c_{i+\delta-\delta'}\rangle + \langle c_{i+\delta-\delta'}^{\dagger} c_i^{\dagger}\rangle \langle c_i c_{i+\delta-\delta'}\rangle$.

For the natural closure matrix $\bf{M}$, we obtain the explicit expression as
$\mathbf{M}_{11}=\text{diag} \left( \epsilon_{\vec{k}}-\mu, \, -\epsilon_{\vec{k}}-\mu, \, -\epsilon_{\vec{k}}+\mu, \, \epsilon_{\vec{k}}+\mu \right)$, $\mathbf{M}_{12}= \text{diag} \left( 2\epsilon_{\vec{k}}, \, -2\epsilon_{\vec{k}}, \, -2\epsilon_{\vec{k}},\,  2\epsilon_{\vec{k}} \right)$,
$\mathbf{M}_{21}= \text{diag}(V,\, V, \,-V,\, -V)$,
and
$\mathbf{M}_{22}= \text{diag}(4V-\mu,\, 4V-\mu, \,-4V+\mu,\, -4V+\mu)$.

\end{subsection}

\begin{subsection}{p-5-sc basis}

This operator basis reads
\begin{equation}      \label{eq.31}
  \vec{A}=\left( c_{\vec{k}},c_{\vec{\pi}-\vec{k}}, c_{-\vec{k}}^{\dagger} , c_{\vec{k}-\vec{\pi}}^{\dagger}, d_{\vec{k}}(\delta),d_{\vec{\pi}-\vec{k}}(\delta), d_{-\vec{k}}^{\dagger} (\delta), d_{\vec{k}-\vec{\pi}}^{\dagger}(\delta) \right)^T.
\end{equation}
It contains all the higher order operators that appear in the commutator $[c_{\vec{k}}, H]$ separately. In the above equation, each symbol depending on $\delta$ represents four operators, with $\delta$ ranging from $1$ to $z=4$. This basis therefore has a dimension of $20$. Due to the complication of Liouville matrix $\bf{L}$, here we also need to use PPA. The first four operators are regarded as block one and the other $16$ operators as block two.

The block matrices of $\bf {I}$ are obtained as follows. $\mathbf{I}_{11}= \text{diag}(1,1,1,1)$, $\mathbf{I}_{21}=(\mathbf{I}_{12})^{\dagger}$,
\begin{equation}       \label{eq:32}
\mathbf{I}_{12} = \left(
\begin{array} {cccc}
    n-\alpha_{\vec{k}}^-(\delta)   & 0   & [\beta_{\vec{k}}^+(\delta)]^{\ast}   & 0  \\
     0    & n+\alpha_{\vec{k}}^+(\delta)     & 0   & -[\beta_{\vec{k}}^-(\delta)]^{\ast}  \\
    \beta_{\vec{k}}^-(\delta)   & 0    & n-[\alpha_{\vec{k}}^+(\delta)]^{\ast}   & 0  \\
    0    & -\beta_{\vec{k}}^+(\delta)    & 0   & n+[\alpha_{\vec{k}}^-(\delta)]^{\ast}  
\end{array} \right),
\end{equation}
and
\begin{equation}       \label{eq:33}
\mathbf{I}_{22} = \left(
\begin{array} {cccc}
    \zeta_{\vec{k}}(\delta,\delta') & 0   & -\eta_{\vec{k}}(\delta',\delta)   & 0  \\
     0    &  \zeta_{\vec{\pi}-\vec{k}}(\delta,\delta')    & 0   & \eta_{\vec{k}}(\delta,\delta')  \\
    -\eta_{\vec{k}}^{\ast}(\delta,\delta')   & 0    & \zeta_{-\vec{k}}^{\ast}(\delta,\delta')   & 0  \\
    0    & \eta_{\vec{k}}^{\ast}(\delta',\delta)    & 0   & \zeta_{\vec{k}-\vec{\pi}}^{\ast}(\delta,\delta') 
\end{array} \right).
\end{equation}
In the above two equations, $\delta$ and $\delta'$ take values $1 \sim 4$. Equations.($\ref{eq:32}$) and ($\ref{eq:33}$) are therefore compact expression for the $4 \times 16$ matrix ${\bf I}_{12}$ and the $16 \times 16$ matrix ${\bf I}_{22}$, respectively. The matrix elements are expressed as the following. 
\begin{align} \label{eq:34}
		& \alpha_{\vec{k}}^{\pm}(\delta)=\frac{1}{N}\sum_{\vec{k}'} e^{i(\vec{k}'\pm \vec{k})\cdot\vec{\delta}}\langle n_{\vec{k}'}\rangle,\nonumber \\
		& \beta_{\vec{k}}^{\pm}(\delta)=\frac{1}{N}\sum_{\vec{k}'} e^{i(\vec{k}'\pm \vec{k})\cdot\vec{\delta}}\langle c_{-\vec{k}'}c_{\vec{k}'}\rangle,\nonumber \\
		& \gamma_{\vec{k}}(\delta,\delta')=\frac{1}{N}\sum_{\vec{k}'} e^{i(\vec{k}'- \vec{k})\cdot\vec{\delta}'}\langle d_{\vec{k}'}^{\dagger}(\delta)c_{\vec{k}'}\rangle,\nonumber \\
		& \eta_{\vec{k}}(\delta,\delta')  \nonumber \\
		& =\frac{1}{N}\sum_{\vec{k}'} \left[e^{i(\vec{k}'+ \vec{k})\cdot\vec{\delta}'}\langle d_{\vec{k}'}^{\dagger}(\delta)c_{-\vec{k}'}^{\dagger}\rangle+
		e^{i(\vec{k}'- \vec{k})\cdot\vec{\delta}} \langle d_{\vec{k}'}^{\dagger}(\delta')c_{-\vec{k}'}^{\dagger}\rangle \right],\nonumber \\
		& \zeta_{\vec{k}}(\delta,\delta')  \nonumber \\
		&=\frac{1}{N}\sum_i\langle n_i n_{i+\delta-\delta'}\rangle-\gamma_{\vec{k}}(\delta,\delta')
		-\gamma_{\vec{k}}^{\ast}(\delta',\delta) -\delta_{\delta+\delta',0}
		[\alpha_{\vec{k}}^-(\delta)]^{\ast}.
\end{align}
%
The natural closure matrix $\mathbf{M}$ can be extracted from Eq.($\ref{eq:25}$) and its Hermitian conjugate. Finally, when $\delta\neq\delta'$, we make an additional approximation to the average $\langle n_{i}n_{i+\delta-\delta'}\rangle$ in $\zeta_{\vec{k}}(\delta,\delta')$, taking $\langle n_{i}n_{i+\delta-\delta'}\rangle \approx \langle n_{i}\rangle \langle n_{i+\delta-\delta'}\rangle - 
\langle c_{i+\delta-\delta'}^{\dagger}c_{i}\rangle \langle c_{i}^{\dagger}c_{i+\delta-\delta'}\rangle + \langle c_{i+\delta-\delta'}^{\dagger} c_i^{\dagger}\rangle \langle c_i c_{i+\delta-\delta'}\rangle$.

\end{subsection}
\end{section}

\begin{section}{HFB mean-field approximation}

For HFB basis, we find that EOM PTA is equivalent to HFB mean-field approximation. In this section, we summarise the analytical formula obtained. From HFB mean-field approximation, Hamiltonian ($\ref{eq:1}$) is reduced to 
\begin{align}\label{eq:35}
&	H_{MF} = \sum_{\vec{k}}\tilde{\epsilon}_{\vec{k}}c_{\vec{k}}^{\dagger}c_{\vec{k}}
-\sum_{\vec{k}} \left[ \Delta (\vec{k} ) c_{\vec{k}}^{\dagger}c_{-\vec{k}}^{\dagger}
+H.c. \right]  \nonumber \\ 
& - \sum_{\vec{k},\vec{k}'}\left[ \frac{2V}{N}-V \left( \vec{k},\vec{k}' \right) \right] \langle n_{\vec{k}}\rangle \langle n_{\vec{k}'}\rangle + \sum_{\vec{k}}\Delta (\vec{k} )
\langle c_{\vec{k}}^{\dagger}c_{-\vec{k}}^{\dagger}\rangle.
\end{align}
In this equation, 
\begin{align}\label{eq:36}
&	\tilde{\epsilon}_{\vec{k}}=	\epsilon_{\vec{k}}+4nV-2\sum_{\vec{k}'}V ( \vec{k},\vec{k}' )\langle n_{\vec{k}'}\rangle - \mu , \nonumber \\ 
& V\left( \vec{k},\vec{k}' \right)=\frac{V}{N}\Big[\cos (k_x-k_{x}^{'} )+ \cos ( k_y-k_{y}^{'} )\Big], \nonumber \\ 
& \Delta(\vec{k})=-\sum_{\vec{k}'} V (\vec{k},\vec{k}' ) \langle c_{-\vec{k}'}c_{\vec{k}'}\rangle\nonumber \\ 
& \hspace{2.1em} = \Delta_x \sin k_x + \Delta_y  \sin k_y
\end{align}
are the renormalized single-particle dispersion relation, the interaction potential matrix, and the energy gap function, respectively. In Eq.($\ref{eq:36}$), $\Delta_x = -(V/N) \sum_{\vec{k}'} \sin (k_x')
  \langle c_{-\vec{k}'}c_{\vec{k}'}\rangle$ and $\Delta_y=- (V/N) \sum_{\vec{k}'} \sin( 
k_y')  \langle c_{-\vec{k}'}c_{\vec{k}'}\rangle$.
From the above equation, we see that $\Delta(\vec{k})$ is an odd function of $\vec{k}$, implying odd parity for the possible superconducting pairing. The last two terms in $H_{MF}$ are constants that only shift the energy.

Solving $H_{MF}$ by Bogoliubov transformation, we obtain the single particle dispersion $\xi_{\vec{k}}=\sqrt{\tilde{\epsilon}_{\vec{k}}^{2}+4
\Delta(\vec{k} ) \Delta^{\ast}(\vec{k} ) }$. The self-consistent equation for the gap function reads
\begin{align}\label{eq:37}
\Delta(\vec{k})=-\sum_{\vec{k}'}V \left(\vec{k},\vec{k}' \right)
\frac{\Delta(\vec{k}')}{\xi_{\vec{k}'}} {\rm tanh}\Big(\frac{\beta\xi_{\vec{k}'}}{2}\Big).
\end{align}
The averages $\langle n_{\vec{k}}\rangle$ appearing in $\tilde{\epsilon}_{\vec{k}}$ is given by
\begin{equation}\label{eq:37p}
\langle c_{\vec{k}}^{\dagger}c_{\vec{k}}\rangle=\frac{u_{\vec{k}}^{2}}{e^{\beta\xi_{\vec{k}} }+1}+\frac{v_{\vec{k}}^{2}}{e^{-\beta\xi_{\vec{k}} }+1},
\end{equation}
with
\begin{align} \label{eq:38}
&	u_{\vec{k}}^{2}=\frac{1}{2}\Big(1+\frac{\tilde{\epsilon}_{\vec{k}}}{\xi_{\vec{k}}}\Big),\hspace{2em}
v_{\vec{k}}^{2}=\frac{1}{2}\Big(1-\frac{\tilde{\epsilon}_{\vec{k}}}{\xi_{\vec{k}}}\Big).
\end{align}
The superconducting transition temperature $T_c$ satisfies the two equations below,
\begin{align}\label{eq:39}
&	1=-\frac{V}{N}\sum_{\vec{k}'}\frac{ {\sin}^2\left( k_{x}^{'} \right)}{\tilde{\epsilon}_{\vec{k}'}} {\rm tanh}\Big(\frac{\beta_c\tilde{\epsilon}_{\vec{k}'}}{2}\Big), \nonumber \\ 
& 1=-\frac{V}{N}\sum_{\vec{k}'}\frac{ {\sin}^2 \left( k_{y}^{'} \right)}{\tilde{\epsilon}_{\vec{k}'}} {\rm tanh}\Big(\frac{\beta_c\tilde{\epsilon}_{\vec{k}'}}{2}\Big).
\end{align}
Due to the $C_4$ symmetry of the lattice, the above two equations in Eq.($\ref{eq:39}$) give the same $T_c$. Obviously, HFB approximation predicts that there is no superconducting phase in the SF model with repulsive interaction $V>0$.

\end{section}

\begin{section}{results}

Using the formalism developed in previous sections, we obtain numerical results for bases HF, p-5, HFB, p-2-sc, and p-5-sc. The PH symmetric formalism is used in all our calculations. We find that the stability of SC is sensitive to lattice size. For HF and HFB bases in momentum space, results are obtained for lattice as large as  $10^3 \times 10^3$ sites. For p-5, p-2-sc, and p-5-sc bases in momentum space, our study is limited to $10^2 \times 10^2$ system. In real space, we can study systems of size $30 \times 30$ using HF, p-5, and HFB. The finite size effects for these sizes are negligible unless stated otherwise.

We first briefly review some basic properties of Hamiltonian ($\ref{eq:1}$) at the PH symmetric parameter $\mu=2V$. In the weak-coupling limit ($V=0$), as pointed out by Gubernatis $et$  $al$.$\cite{JG31}$, the Fermi surface has a nesting momentum ($\pi,\pi$) and the single-particle density of states (DOS) has logarithmic Van Hove singularity at Fermi energy. At low temperature, they cause the divergence of density-density susceptibility at ($\pi,\pi$) and ($0,0$), respectively. As a consequence, the density instability prefers to appear at $\vec{k}$ = ($\pi,\pi$) (for $V>0$) and $\vec{k}$ = ($0,0$)   (for $V<0$).

The strong-coupling perturbation analysis$\cite{JG31}$ up to $(t/V)^2$-order shows that Eq.($\ref{eq:1}$) can be mapped into a two-dimensional XXZ model in the z-directional magnetic field. In the limit $\vert V\vert \to \infty$, SF model becomes an Ising model. The $Z_2$ symmetry of the Ising model at zero field corresponds to the PH symmetry of SF model at half filling. This symmetry can be spontaneously broken at $T < T_c=0.56\vert V\vert$. For $V<0$ (ferromagnetic Ising coupling), this spontaneous symmetry breaking gives out a ferromagnetic phase for Ising model, or a phase separation for SF model. The $U(1)$ symmetry of SF model could be further broken at a lower temperature. For $V>0$ (anti-ferromagnetic Ising coupling), the breaking of $Z_2$ symmetry is accompanied with the breaking of sublattice translation symmetry, giving an anti-ferromagnetic phase, or the CO phase for SF model. 

We therefore conclude that the PH symmetry at $\mu=2V$ could be broken independently (leading to phase separation of normal states) or together with other symmetries of the SF model, such as the A-B lattice translation symmetry (leading to charge ordering order) or $U(1)$ symmetry (leading to phases separated superconducting states). As we will see below, a unified phase diagram discloses the interplay of these symmetry breakings at the thermodynamical level.

\begin{figure}[htb]
\vspace{-0.0cm}
\begin{center}
\includegraphics[width=3.40in, height=2.80in, angle=0]{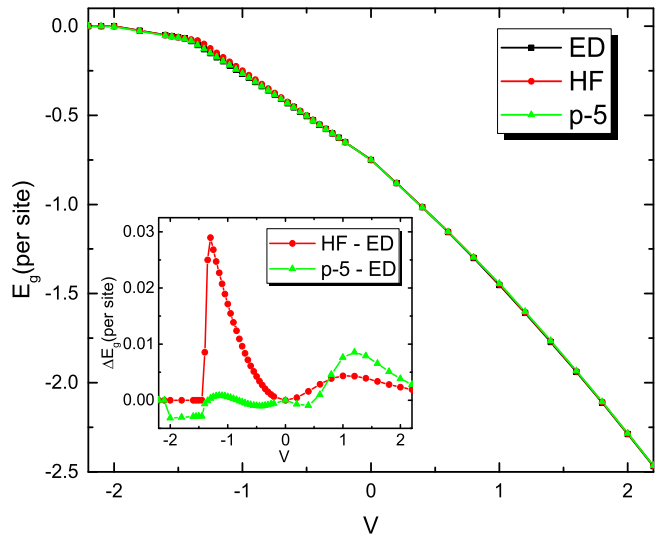}
\vspace{-0.80cm}
\end{center}
  \caption {Ground state energy per site for $4 \times 4$ lattice from ED, HF, and p-5 basis. Parameters are $T=0.0$ and $\mu=2V$. (Inset) Ground state energies obtained by HF (dots) and p-5 basis (upper triangles) subtracting that from ED. The lines are for guiding eyes.}\label{Fig1}
\end{figure}

In Fig. $\ref{Fig1}$, we plot the ground state energy per site as functions of $V$ obtained from HF and p-5 basis for a $4 \times 4$ lattice. They are compared to the exact energy obtained from exact diagonalization (ED). The fermion density is fixed at half filling, i.e., $\mu = 2V$ is used. The three curves are quite close to each other on the scale of the main figure. A weak change of slope can be observed at $V=0$, which corresponds to transition into different phases in the thermodynamical limit. As to be shown in Fig.$\ref{Fig3}$ and Fig.$\ref{Fig7}$ for infinitely large system, in the regime $V > 0$, the sublattice translation symmetry of the square lattice is broken and the ground state is in CO phase. For $V<0$, there are two degenerate ground states, one with high density ($n_h=0.5+\delta$) and another with low density ($n_l=0.5-\delta$), both in superconducting phase. If inhomogeneity were allowed in the solution, they would coexist in real space and occupy equal volume of the sample to give a nominal filling of $1/2$.

The abrupt change of slope at $V=-1.4$ in the curves is a finite size effect that arises from the level crossing of ground state energies between the $n=1/16$ (in $V < -1.4)$ and $n=5/16$ (in $V > -1.4$) subspaces. Due to PH symmetry, another level crossing occurs at same $V$ between $n=15/16$ (in $V < -1.4)$ and $n=11/16$ (in $V > -1.4$) subspaces.

The Inset of Fig.$\ref{Fig1}$ shows the error of ground state energy per site obtained by HF and p-5 basis, taking ED result as reference. HF always gives non-negative error, reflecting the variational nature of HF approximation. For $V<-1.4$, the energy error from HF is zero because it correctly describes that the ground states has only one particle or hole and there is no correlation. Another point of zero correlation is at $V=0$, where HF approximation is exact. The largest error of HF curve occurs at $V \gtrsim -1.4$ due to large charge fluctuations, reaching a relative error $\left[E_g(HF)-E_g(ED)\right]/ E_g(ED) \sim 20\%$.
In contrast, the energy error from p-5 basis is not variational. It is much smaller than HF result at $V \gtrsim -1.4$. In the regime $-2.0 < V < -1.4$, p-5 basis gives inaccurate fermion densities and it leads to a relative energy error $\sim 10\%$. For $V < - 2.1$, the energy from p-5 basis becomes accurate again. At $V=0$, exact energy is obtained by p-5. For large repulsive interaction, both HF and p-5 give small energy errors due to frozen of charge fluctuations in this regime, with relative errors less than $0.3\%$ (HF) and $0.6\%$ (p-5), respectively.
This comparison of energy errors show that while p-5 is not variational, it gives overall smaller energy errors in the regime $-1.4 < V < 0.8$ where charge fluctuations are large. This reflects that correlations are taken into account by larger operator basis. Note that for this small lattice, p-5-sc basis produces zero SC order parameter and the energy is the same as that from p-5 basis.

\begin{subsection}{attractive interaction: phase separation and superconductivity}

\begin{figure}[htb]
\vspace{-0.0cm}
\begin{center}
\includegraphics[width=3.40in, height=3.0in, angle=0]
{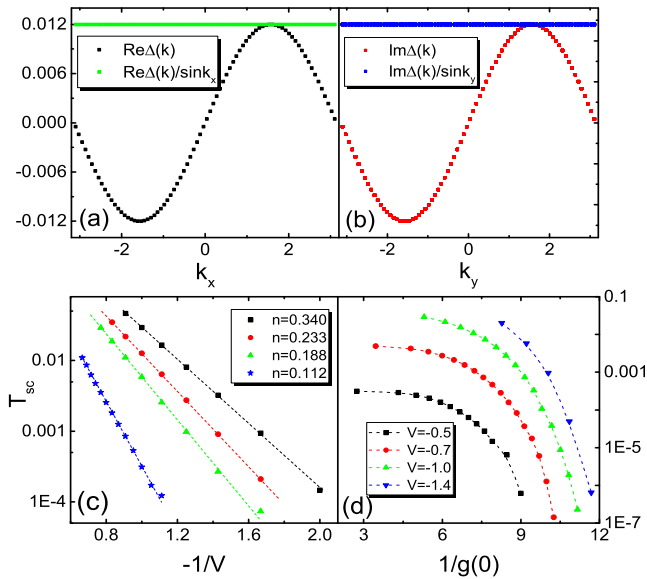}
\vspace{-0.80cm}
\end{center}
\caption {HFB results for uniform SC phase. (a) $\text{Re}\Delta(\vec{k})$ as a function of $k_x$. (b) $\text{Im}\Delta(\vec{k})$ as a function of $k_y$. Parameters are $T=0.0$, $V=-1.0$, and $\mu=-2.2$. (c) Superconducting transition temperature $T_{sc}$ as a function of interaction $V$ for different fermion density $n$. (d) $T_{sc}$ as a function of $1/g(0)$ for different $V$. $g(0)$ is the density of states at Fermi energy. The dashed lines are for guiding eyes.}\label{Fig2}
\end{figure}
In this section, we focus on the case of $V < 0$. We first discuss the results from HFB basis whose solution can be analysed in more detail due to the analytical formula presented in Sec.IV. 

\subsubsection{HFB result}

We first study the properties of the uniform SC state obtained from HFB basis. For this purpose, we choose parameters such that the particle filling is far away from half filling to guarantee that the solution is in an uniform SC phase. For fillings close to half filling, the uniform SC state is unstable towards PS. The interplay of PS and SC will be discussed using the phase diagram in the next subsection. The results below are obtained by solving the analytical self-consistent equations in Sec.IV.

We find that the gap function $\Delta(\vec{k})$ is complex. $\text{Re} \Delta(\vec{k})$ and $\text{Im} \Delta(\vec{k})$ depends only on $k_x$ and $k_y$, respectively. Figures $\ref{Fig2}$(a) and 2(b) show their curves at $T=0$ obtained from the self-consistent solution of Eq.($\ref{eq:37}$), which agree with the form $\Delta(\vec{k},T)=\Delta(T)({\rm sin}k_x+i \, {\rm sin}k_y)$, i.e., in $p_x + ip_y$ symmetry. This is consistent with the analysis of Cheng $et$ $al$.$\cite{MC37}$ Note that although the pairing average $\langle c_{-\vec{k}}c_{\vec{k}} \rangle$ is sharply distributed around Fermi surface, $\Delta(\vec{k}, T)$ does not depends directly on the shape of Fermi surface. We find that the ground state energy of $p_x + ip_y$-wave uniform SC without nodal line is always lower than that of a $p_x$-wave (or $p_y$-wave) uniform SC with a nodal line (For the inhomogeneous solution, $p_x + ip_y$-wave SC has higher energy than $p_x$-wave, see below). This can be understood since the nodal line has no contribution to the condensation energy. Mathematically, this is related to the fact that the mean-field free energy $F$ is a concave function of the norm of the superconducting order parameter $\Delta$ (i.e. $F''(\vert\Delta\vert^2)>0$).$\cite{MC37}$
The relative difference between the ground state energies of SC and normal phases is found to be less than $10^{-3}$, which leads to a low superconducting phase transition temperature (see below).


The SC critical temperature $T_{sc}$ obtained from Eq.($\ref{eq:39}$) is plotted versus $-1/V$ for various fermion densities in Fig.$\ref{Fig2}$(c). For each curve, the chemical potential is tuned to keep $n$ fixed. Figure $\ref{Fig2}$(c) supports the exponential dependence $T_{sc}\sim \text{exp}\left[-1/( \alpha\vert V\vert ) \right]$, which is different from $T_{tp}\approxeq 2 \text{exp}(-2\pi/\sqrt{\vert V\vert})$ obtained by random phase approximation (RPA)$\cite{JG31}$ at half filling in the weak attraction regime $-0.38 < V < 0$. For $V<-0.38$, RPA predicts that the PS transition temperature $T_{ps}$ exceeds $T_{sc}$ and the superconducting pairing will be suppressed by PS at low temperatures. The difference traces back to the fact that Gubernatis $et$ $al$. considered an uniform SC at half filling where the van Hove singularity on the Fermi surface enhances the SC.$\cite{JG31}$ Our calculation at the nominal filling $n=1/2$, in contrast, obtains a PS between SC states with two actual fillings $n_{l} < 1/2$ and $n_{h} > 1/2$, for each of which the van Hove singularity lies away from Fermi surface and does not influence $T_{sc}$. Due to this PS at half filling, an homogeneous SC is thermodynamically unstable at $n = 1/2$, as to be discussed in Fig.$\ref{Fig3}$. Figure $\ref{Fig2}$(c) also shows that for a fixed $V$, $T_{sc}$ increases dramatically with increasing density. This can be largely attributed to the increase of density of states at Fermi energy $g(0)$ with increasing $n$.
Due to the PH symmetry of the Hamiltonian, on the high density side $n>1/2$, $T_{sc}$ will decrease as $n$ increases.

Figure $\ref{Fig2}$(d) shows the dependence of $T_{sc}$ on $1/g(0)$ for various $V$ values. Here, the single-particle density of state at Fermi energy $g(0)$ is calculated from $g(\omega)=(1/N)\sum_{\vec{k}}\delta(\omega-\tilde{\epsilon}_{\vec{k}})$ with Lorentz broadening of $\delta$ functions. It changes with the filling of fermions, which is in turn tuned by $\mu$. The curves in Fig.$\ref{Fig2}$(d) deviate significantly from exponential form, in contrast to that in the BCS superconductivity. This is because the attractive interaction $V$ in our model Hamiltonian ($\ref{eq:35}$) is not limited to the Debye shell around Fermi surface. Accordingly, in Eq.($\ref{eq:39}$), the sum of momentum runs over the entire first Brillouin zone rather than within the Debye energy shell around the Fermi surface.

\subsubsection{Results from larger bases}

\begin{figure}[t!]
\vspace{-0.0cm}
\begin{center}
\includegraphics[width=3.40in, height=2.80in, angle=0]{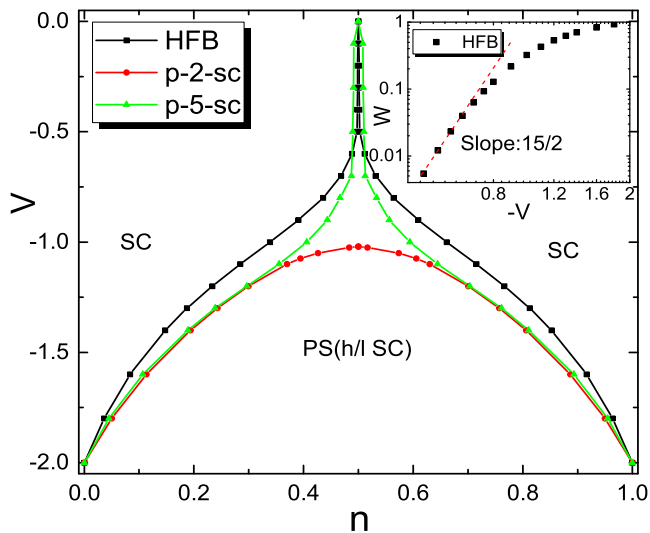}
\vspace{-0.80cm}
\end{center}
  \caption {Ground state phase diagram on $V$-$n$ ($V<0$) plane. SC: superconducting phase; PS(h/l SC): phase separation between high- and low-density superconducting phases. Inset: the width of PS region $W =n_h-n_l$ as a function of $-V$ obtained from HFB basis. The dashed line is a fitting with slope $7.5$.}\label{Fig3}
\end{figure}

The zero temperature $V-n$ phase diagram (on the half plane of $V<0$) is presented in Fig.$\ref{Fig3}$. The phase boundaries  obtained from HFB and p-5-sc are qualitatively similar. They divide the diagram into three regions, two SC phases in the low and the high density regimes respectively, and PS between them near $n=1/2$. The higher boundary density $n_h$ is obtained by scanning chemical potential upwards and observe that fermion density jumps at certain $\mu$ to a larger value $n_h$. The lower boundary density $n_l$ is obtained similarly from the inverse scanning. This approach gives slightly wider density window of PS than the binodal lines obtained from Maxwell construction based on the $S$-shape $n$-$\mu$ curve.$\cite{NHT38.1,NHT38.2,NHT38.3}$ Fig.$\ref{Fig3}$ implies that close to $n=1/2$, two SC states with different fermion densities coexist in real space, with volume fractions determined by the boundary values $n_l$, $n_h$ and nominal density $n$. Each of the coexisting SC states has the properties of an homogeneous SC phase at the same filling. In reality, long-range interactions beyond our model Hamiltonian may lead to domains or other inhomogeneous structures in the sample.

In Fig.$\ref{Fig3}$, the PS region obtained from HFB, p-2-sc, and  p-5-sc bases are qualitatively the same when $|V|$ is large. In the small $|V|$ region, the results of HFB and p-5-sc are qualitatively different from those of p-2-sc. 
The phase boundaries from HFB and p-5-sc have a sharp peak around $n=1/2$ and in small $\vert V\vert$, while there is no PS in p-2-sc in this region (say $|V|<1.0$).
 Quantitative comparison shows that HFB basis gives a sharper peak of PS region at $n=1/2$ and $V > -0.5$ than p-5-sc, while p-5-sc gives almost identical boundary for $V < -1.1$ as p-2-sc. For HFB and p-5-sc bases, the width of PS region increases with increasing $|V|$ and decreases to zero only at $V=0$. 
That PS occurs at $n=1/2$ for any finite attractive $V$ is consistent with the notion that the PS discussed here is a density instability due to divergence of density susceptibility at momentum $(0,0)$, which is in turn caused by the van Hove singularity at $n=1/2$.$\cite{JG31}$
Therefore, we speculate that the disappearance of PS for the p-2-sc basis in small $|V|$ region is non-physical, which may be due to the improper estimation of the relative weight between operators $c_{\vec{k}}$ and $d_{\vec{k}}$ in the p-2-sc basis by inner product Eq.($\ref{eq:6}$).

The inset of Fig.$\ref{Fig3}$ shows that the width of PS region $W$ (the density difference between high- and low-density SC)
in the thermodynamic limit decreases as a power law $W\sim \vert V\vert^{7.5}$. It is obtained from HFB basis for which calculation can be done for the number of lattice sites as large as $N \sim 10^6$ in momentum space. The power law behavior is consistent with the singular nature of the point ($V=0$, $n=1/2$) on the phase diagram and may be related to the van Hove singularity. A complete understanding of it is still absent.
For p-5-sc basis, our computation is limited to lattice sites $N \sim 10^4$ in momentum space. The obtained width of PS region in Fig.3 (between two green lines) is wider and does not follow power law in $|V|=0$ limit. This is due to the finite size effect since we observed that with increasing size, the width obtained from p-5-sc basis decreases. 

\begin{figure}[t!]
\vspace{-0.0cm}
\begin{center}
\includegraphics[width=3.40in, height=2.80in, angle=0]{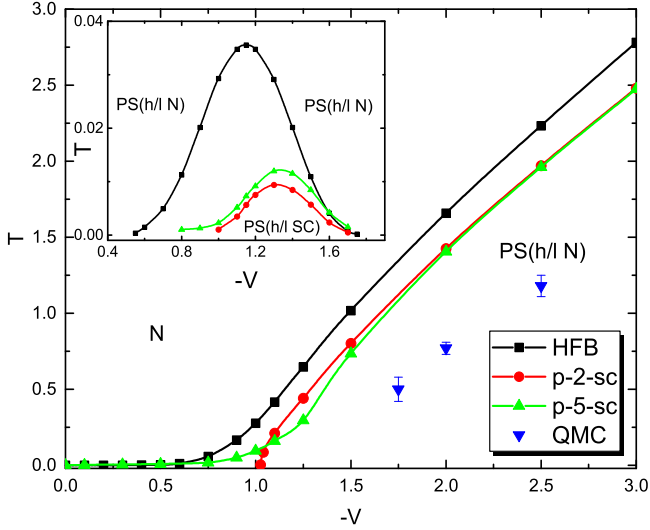}
\vspace{-0.80cm}
\end{center}
  \caption {Phase diagram on $T$-$|V|$ ($V<0$) plane at half filling $\mu=2V$. The symbols with guiding lines in the main figure are the PS transition temperature $T_{ps}$. Thoses in the inset are the SC transition temperature $T_{sc}$. N: homogeneous normal phase; PS(h/l N): coexistence of high-density and low-density homogeneous normal phases; PS(h/l SC): coexistence of high-density and low-density superconducting phases. The quantum Monte Carlo (QMC) results are from Gubernatis et al.$\cite{JG31}$ } \label{Fig4}
\end{figure}

Figure $\ref{Fig4}$ shows the phase diagram on the $T$-$|V|$ ($V<0$) plane at half filling. Symbols with eye-guiding lines in the main figure mark the PS transition temperature $T_{ps}$ obtained from different bases of EOM PTA. At high temperatures, the system is in a homogeneous normal phase. For $T$ below $T_{ps}$, PS occurs and two normal phases with different densities $n_{h}$ and $n_{l}$ coexist (PS(h/l N) in the main figure). The appearance of PS means that the PH symmetry of Hamiltonian is spontaneously broken. In the correspondence of SF and Ising model in large $V$ limit, PS between high-/low-density normal phases at low temperature corresponds to the magnetized phases of spin-up/spin-down. 

The inset of Fig.$\ref{Fig4}$ shows the SC transition of phase-separated states at a much lower temperature $T_{sc}$ ($T_{sc} \sim 10^{-2} \ll T_{ps}$). At the common $T_{sc}$, each of the coexisting normal phases undergoes the SC transition, making a state of SC-SC coexistence below $T_{sc}$ (PS(h/l SC) in the inset of Fig.$\ref{Fig4}$). The maximum value of $T_{sc}$ is obtained at $V=-1.2$ for HFB and at $V=-1.3$ for p-2-sc and p-5-sc.
Due to the huge difference in magnitude ($T_{sc} \sim 10^{-2} T_{ps}$ in the whole negative $V$ regime), we plot the SC phase boundary separately in the inset. Combined together, the two figures give the scenario that for $V<0$ and half filling, as temperature decreases, the state first transits from an homogeneous normal state into a phase separated normal state at $T_{ps}$, and then transits into phase separated SC state at $T_{sc}$, showing successive breaking of PH symmetry and $U(1)$ symmetry with decreasing temperature.

Now we compare different $T_{ps}-V$ curves in the main figure of Fig.$\ref{Fig4}$. The results of HFB and p-5-sc are closer when $\vert V\vert$ is small, and those from p-2-sc and p-5-sc are closer when $\vert V\vert$ is large. The $T_{ps}$'s decrease rapidly around $V=-1.0$. In particular, p-2-sc basis produces $T_{ps}=0$ for $V<-1.0$. These observations are consistent with the results in Fig.$\ref{Fig3}$. $T_{ps}$ from p-2-sc and p-5-sc bases scale as $T_{ps}\simeq 0.9\vert V\vert$ in the large $|V|$ limit. They are improved with respect to the HFB result $T_{ps}\simeq \vert V\vert$, but are still much higher than QMC data (down triangles) and the exact behavior $T_{ps}\simeq 0.56\vert V\vert$.

\begin{figure}[t!]
\vspace{-0.0cm}
\begin{center}
\includegraphics[width=3.40in, height=2.80in, angle=0]{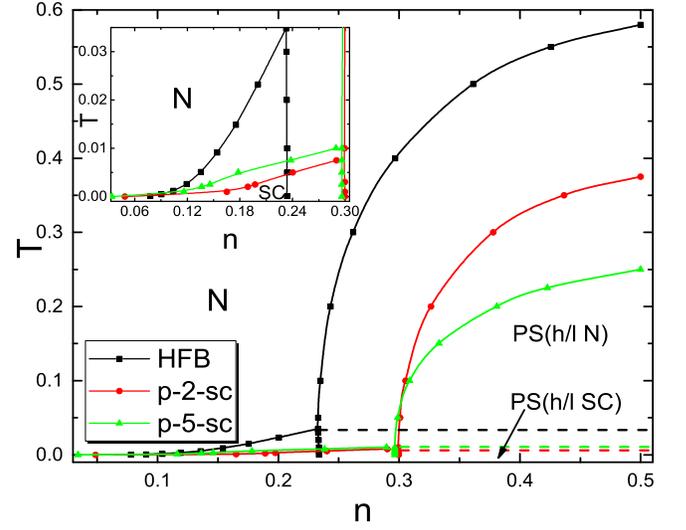}
\vspace{-0.80cm}
\end{center}
  \caption {Phase diagram on $T$-$n$ plane at $V=-1.2$. (Inset) Enlarged	figure in the small $n$ and low-temperature regime. N: hormogeneous normal phase; SC: homogeneous superconducting phase; PS(h/l N): coexistence of high- and low-density normal phases. PS(h/l SC): coexistence of high- and low-density SC phases.}\label{Fig5}
\end{figure}

Figure $\ref{Fig5}$ presents the $T$-$n$ phase diagram at a generic attraction $V=-1.2$. Only $n \leq 1/2$ region is shown since the phase diagram is symmetric with respect to $n=1/2$. Different bases give qualitatively similar phase diagram. Note that in the inset, $T_{sc}$ from p-5-sc is slightly higher than that from p-2-sc. The PS transition temperature $T_{ps}$ has a dome shape with the highest value at $n=1/2$. For a fixed nominal $n$ inside this dome and at high temperature, PS occurs between two normal phases with high density $n_h$ and low density $n_l$. As temperature decreases, the coexisting normal phases will transit into coexisting SC phases below $T_{sc}$ (dashed lines). Due to the PH symmetry of the system, $n_h + n_l=1$ always holds in this process. For $n$ outside this dome (far away from half filling), the system is in an homogeneous normal phase (N) for $T>T_{sc}$ and transits to an homogeneous SC phase (SC) below $T_{sc}(n)$ which has a long tail extending to $n=0$ (see inset for the enlarge figure).
Note that the homogeneous Bogoliubov approximation$\cite{MC37}$ produces an homogeneous SC at $n=1/2$, while this work produces a coexisting high-/low-density SC phase. The existence of PS suppresses $T_{sc}$ because the SC state only occurs at densities far away from $n=1/2$ which has lower $T_{sc}$. In this sense, PS and SC competes at the thermodynamics level. 

\begin{figure}[t!]
\vspace{-0.40cm}
\begin{center}
\includegraphics[width=3.40in, height=3.0in, angle=0]{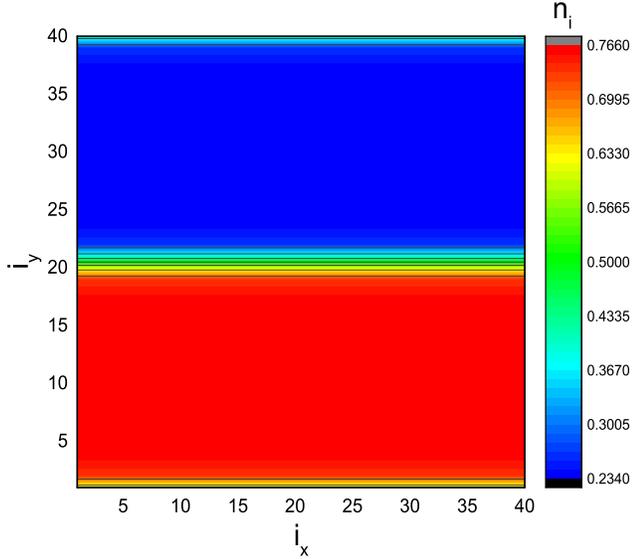}
\vspace{-0.80cm}
\end{center}
  \caption {Real space distribution of density $\langle n_i\rangle$ for $V=-1.2$, $\mu=2V$ and $T=0$, obtained from real space HFB basis on a $40 \times 40$ lattice with periodic boundary condition. The ground state is found to be in SC order with $p_x$ wave symmetry. }\label{Fig6}
\end{figure}

Using the real space HFB basis, we also studied the inhomogeneous SC state in the PS regime without translation symmetry. Figure $\ref{Fig6}$ shows the ground state fermion density distribution $n_i$ for $V=-1.2$ at particle-hole symmetric point $\mu =2V$ on a $40 \times 40$ lattice with periodic boundary. In the calculation, $n_i$ for each site $i$ is allowed to change self-consistently, starting from arbitrary initial conditions. Figure $\ref{Fig6}$ shows one of the stable state obtained. A domain wall lies parallel to $x$ axis and separates the sample into high- and low-density domains with equal volume, making the whole system at nominal half filling. The domain wall is composed of approximately half-filled sites and has a width of several lattice constants. The whole system is in a non-homogeneous SC state with $p_x$ symmetry, which is in contrast to the $p_x + i p_y$ symmetry obtained from the translation symmetric calculation. Using different initial states for the self-consistent calculations, we can obtain the energy-degenerate $p_y$-wave SC state with a domain wall parallel to $y$ axis, but never stabilize a $p_x + i p_y$ wave state with a domain wall. The energy calculation shows that the ground state energies of homogeneous $p_x$ and $p_x + i p_y$ SC states are very close to each other. The existence of a domain wall across the whole sample could well change the energy difference as well as the symmetry of the SC order parameter. 

We also studied the SC-SC interface energy. For this purpose, let us consider a virtual PS(h/l SC) system without SC-SC interface. The ground state energy of this virtual system is the same as that of high (or low) density homogeneous SC. The SC-SC interface energy is thus defined as the ground state energy of the inhomogeneous PS(h /l SC) state (with interface) minus that of the virtual PS(h /l SC) system (without interface).
We find that the SC-SC interface energy is positive and proportional to the system linear size $L$ ( $L=\sqrt{N}$). This seems reasonable since the existence of an interface limits the motion of spinless fermions and increase the kinetic energy. For the high temperature PS(h/l N) phase, the normal-normal interface energy can be defined similarly and we also find a $L$-linear positive interface energy.

\end{subsection}

\begin{subsection}{Repulsive interaction: charge order and phase separation}

\begin{figure}[t!]
\vspace{0.0cm}
\begin{center}
\includegraphics[width=3.40in, height=2.80in, angle=0]{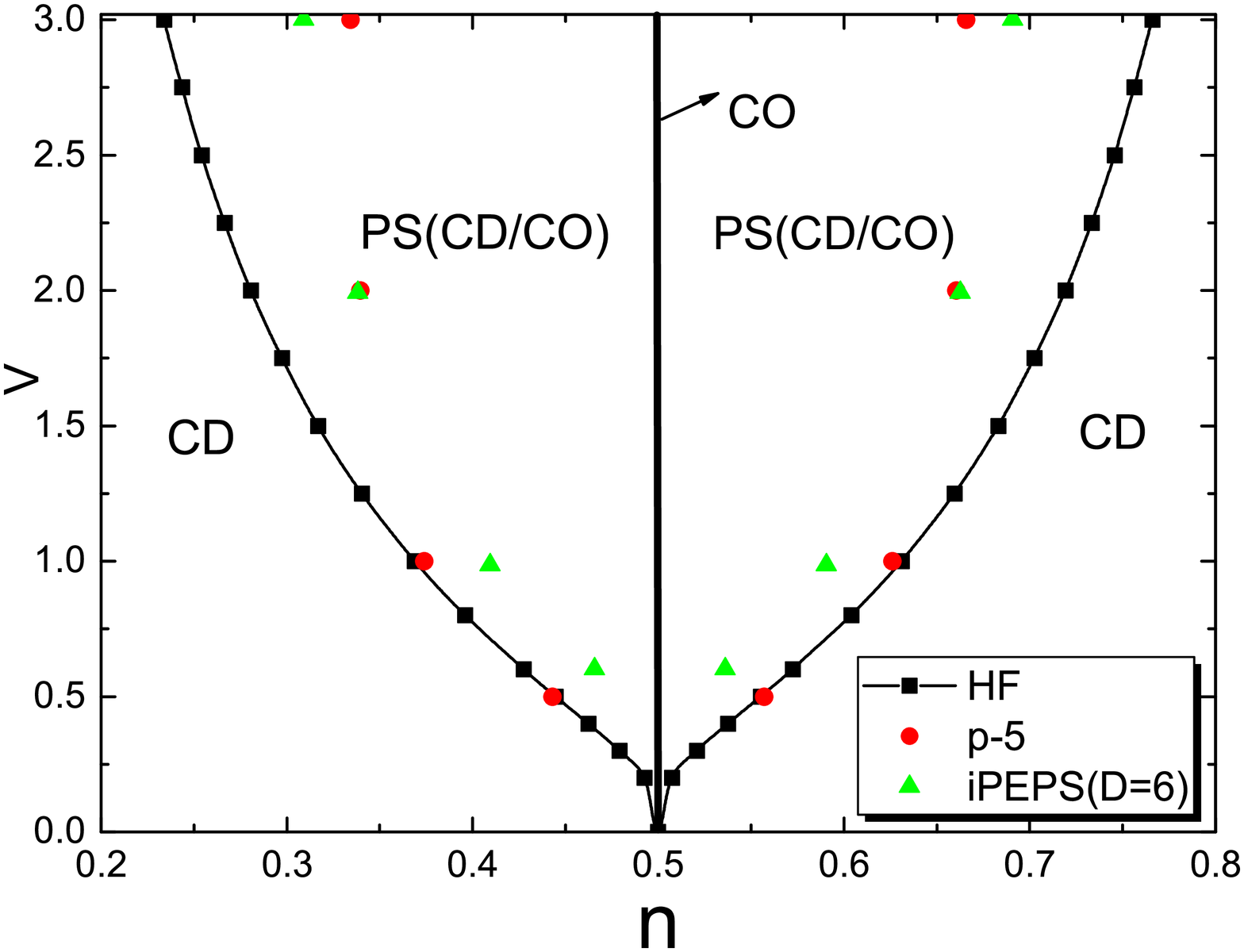}
\vspace{-0.70cm}
\end{center}
  \caption {Ground state phase diagram on $V$-$n$ plane for $V>0$. CD: charge-disordered phase; CO: A-B type charge-ordered phase; PS(CD/CO): coexistence of charge-disordered phase and A-B type CO phase. The result of fermionic projected entangled-pair states (iPEPS) is from Corboz $et$ $al$.$\cite{PC33}$ $D$ is the bond dimension of the iPEPS.}\label{Fig7}
\end{figure}

In this subsection, we study the spinless fermion model with repulsive interaction $V>0$.
Figure $\ref{Fig7}$ shows the ground state $V$-$n$ phase diagram, obtained from HF and p-5 bases. The data from projected entangled-pair states calculation$\cite{PC33}$ are also shown for comparison. The phase boundaries from various calculations are qualitatively consistent. Our calculation gives a CO phase at $n=1/2$ for any finite $V$, as expected from the Fermi surface nesting with nesting momentum $(\pi, \pi)$ and being consistent with the renormalization group analysis.$\cite{RS38}$ This is different from the case of honeycomb lattice where CO occurs only for $V > V_c > 0$.$\cite{LW21}$ For density far away from half filling, an charge-disordered (CD) state is stable. In the regime very close to $n=1/2$, we obtain the PS between CO and CD phase, where CO and homogeneous fermion gas coexist in real space. The width of this coexistence region in the $n$ axis decreases to zero smoothly in the limit $V=0$. In the CO phase, PH symmetry and sublattice translation symmetry are spontaneously broken. Being different from the attractive case, for the repulsive interaction, PS occurs only when holes or particles are doped into CO state and hence it is not a spontaneous PH symmetry breaking.

Comparing the results from HF, p-5 and iPEPS, we find that the agreement is reasonable. In particular, in the small $V$ range, the numerical results of the three methods are relatively close. When the repulsive interaction is strong, the PS boundary obtained by p-5 and iPEPS is more inward.

\begin{figure}[t!]
\vspace{-0.0cm}
\begin{center}
\includegraphics[width=3.40in, height=2.950in, angle=0]{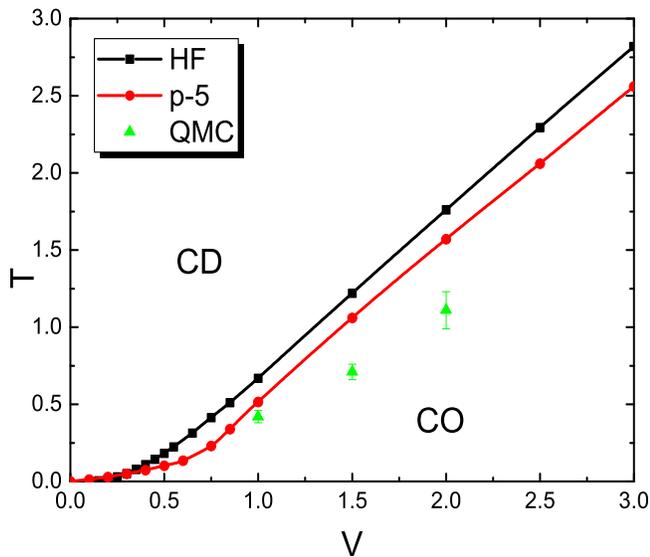}
\vspace{-0.90cm}
\end{center}
  \caption {Charge ordering temperatures $T_{co}$ as functions of repulsive $V$ for half filling $n=1/2$. CD: charge-disordered phase; CO: A-B type charge-ordered phase. The quantum Monte Carlo (QMC) results are from Gubernatis $et$ $al$.$\cite{JG31}$} \label{Fig8}
\end{figure}

Figure $\ref{Fig8}$ shows the critical temperature of CO at half filling as a function of $V$, obtained from HF and p-5 bases. For comparison, QMC data from Gubernatis $et$ $al$.$\cite{JG31}$ are also shown. In the small $V$ limit, HF gives exponentially small $T_{co}$ $\cite{JG31}$ and our p-5 calculation gives consistent results. In the large $V$ limit, p-5 basis gives a linear $T_{co}(V)$ curve with smaller slope than HF result. But both HF and p-5 deviate significantly from the expected Ising results $T_{co} = 0.56V$ in this limit. This reflects that the excitations included in p-5 basis are still insufficient for an accurate description of the thermal excitations of Ising model.

According to Fig.$\ref{Fig8}$, the whole phase diagram on $V$-$n$ plane will change as temperature increases from zero. In Fig.$\ref{Fig9}$, we show such a phase diagram at $T=0.1$. Compared to the zero temperature phase diagram in Fig.$\ref{Fig7}$, CO at $n=1/2$ melts first from the small $V$ regime, recovering the PH and translational symmetry. Accompanying with this melting, the PS between CO and CD state disappears. The CD phases in $n>1/2$ and $n<1/2$ regimes are connected in the small $V$ regime. In the large $V$ regime, CO is no longer limited at $n=1/2$ but extends to a finite density regime around half filling. A second-order charge order-disorder transition line appears near half filling and small $V$.

\begin{figure}[t!]
\vspace{-0.0cm}
\begin{center}
\includegraphics[width=3.40in, height=2.950in, angle=0]{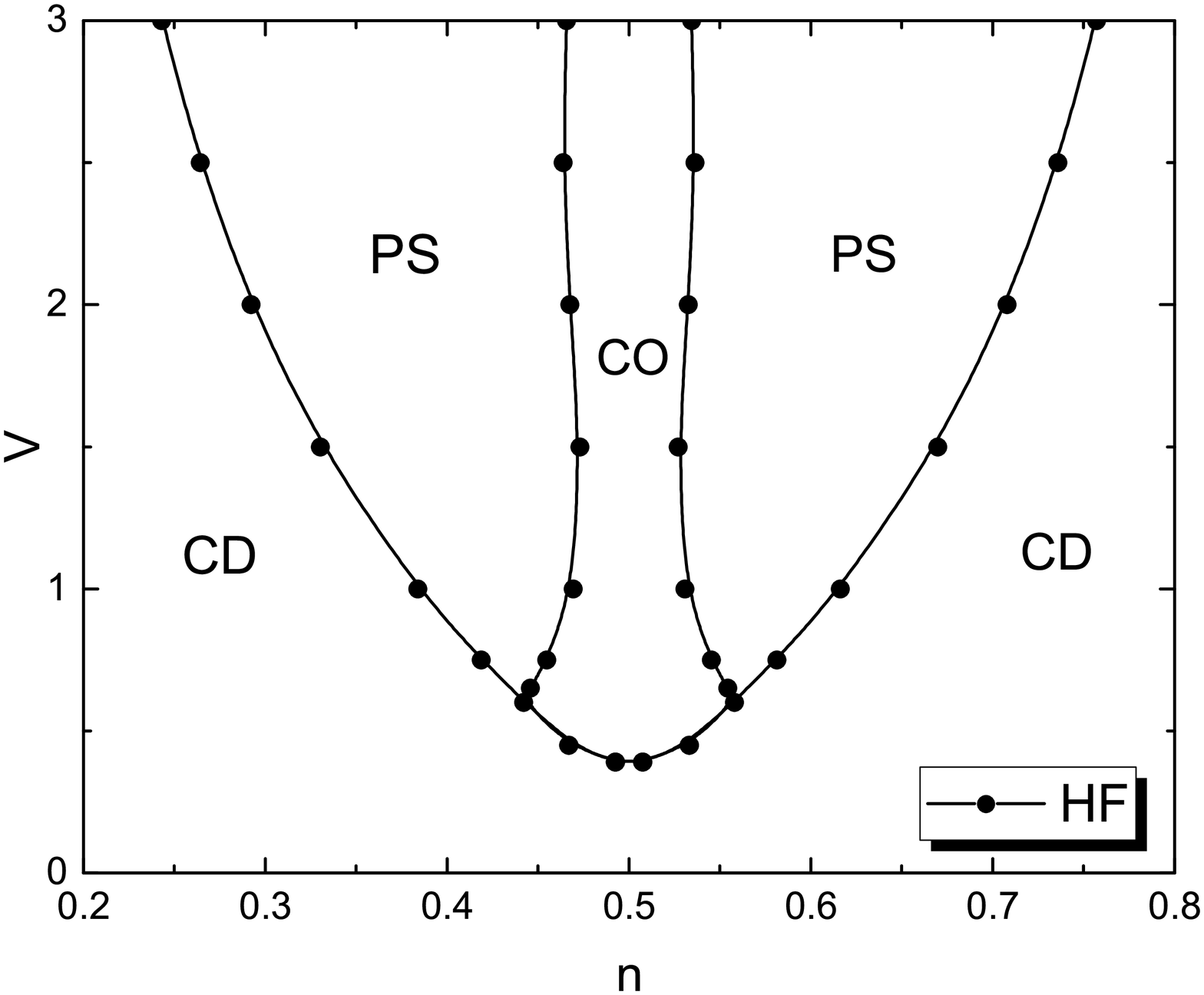}
\vspace{-0.80cm}
\end{center}
  \caption {$V$-$n$ phase diagram for $V>0$ at $T=0.1$ obtained from HF basis. }\label{Fig9}
\end{figure}

\end{subsection}

\end{section}

\begin{section}{summary and discussion}

In summary, in this work, we systematically study the SF model on the square lattice with nearest-neighbor hopping and interaction. 
For the attractive interaction, at low temperature, the system is in the $p+ip$ SC phase when particle concentration $n$ is far away half-filling and in the PS between high- and low- density SC phase near half filling. With the increase of temperature, the homogeneous SC phase will transit into into homogeneous normal phase above $T_{sc}$, while the PS of SC will transit into PS of normal phases which finally transits into homogeneous normal phase at $T>T_{ps}$. Using the real space basis, we observed the domain structure of SC with $p$-wave symmetry at $n=1/2$ and obtain positive interface energy. For repulsive interaction, homogeneous CO state is stable only at half filling. In the low/high fermion density regime, uniform CD phase is stable. In the weakly particle/hole doped regime, a PS between CO and CD phase occurs. Upon increasing temperature, the parameter regime of CO moves to finite $V$ and expands into a finite density regime around half filling.

There are several issues worthy of discussion.
Firstly, in principle, the obtained results can be improved by expanding the basis. In practice, the feasibility of using larger bases depends severely on the strategy of expanding the basis. In this work, we generate operator bases by successively applying the Liouville superoperator	on $c_i$ and collecting each individual operators generated. Starting from $c_i$ (HF basis), $[c_i, H]$ produces the p-5 basis. If we collect all the operators generated by $[ [c_i, H], H]$ into the basis, the dimension of the basis will increase so rapidly that writing down the matrices $I$ and $L$ by hand is already infeasible. We could add only part of the newly generated operators, or certain combinations of new operators (as we did for p-2-sc) into the basis. For example, if only the type $n_{i+\delta}n_{i+\delta'}c_i$ is added to p-5-sc, the computational cost will increase by about $8$ times, which is acceptable. Other ways of expanding the basis include the Lanczos process,\cite{Lee1} or simply collecting all the operators of the form $c_k^{\dagger}c_p c_q$. The former produces continued fraction form of GF but involves correlation functions that are hard to compute. The latter generates a huge basis size of order $N^3$. We estimate that a lattice of $10 \times 10$ sites could be studied with the latter basis using translation symmetry.

 In the expansion of the basis, priority should be given to those operators that describe important fluctuations for the problems under consideration. A systematic and controllable way to expand the basis must take care of both computational complexity and efficiency of the operators. A quantitative criterion for the importance of a basis operator is still lacking. In this sense, it is still an open question how to best extend the operator bases beyond p-5-sc. Maybe the idea of renormalization group, as being successfully adopted by algorithms in Hilbert space diagonalization such as numerical renormalization group and
density matrix renormalization group, could be applied in Liouville space to establish the optimal EOM method in the future.
 
Secondly, the possibility of superconductivity in the repulsive interaction case is an interesting issue. For the system with a sharp Fermi surface and weak repulsive interaction, the effective attraction between particles could be generated through the Kohn-Luttinger mechanism.$\cite{WK39,VMG39}$ However, our PTA always produces CD, CO, and PS(CD/CO) for $V>0$ by the present bases. No SC phase is observed so far from PTA. Considering that PS means that particles prefer to segregate in real space and can be regarded as a signal of effective attractive interaction,$\cite{GU25}$ further study in this direction by expanding the basis is desirable.

Thirdly, the formalism obtained in this paper can be directly extended to the $t-t'-V$ model (that is, taking into account the next-nearest-neighbor hopping $t'$ in Eq.($\ref{eq:1}$)).
In the study of high $T_c$ SC, long-range hopping plays an important role. For example, i) the existence of long-range hopping may better-screening Coulomb repulsion and reproduce the flat band and Fermi surface shape of cuprates; ii) for some unconventional superconductors, $t'$ can not be ignored (e.g., $La_2CuO_4$, $YBa_2Cu_3O_7$, $Bi_2Sr_2CaCu_2O_8$, $LaNiO_2$, etc.) and it influences the SC transition temperature.$\cite{RR39,EP39,KT39,CS39,HJ39,AB39,ST39}$ Moreover, the existence of $t'$ will destroy the PH symmetry of the system and change the present results significantly. Previous HF approximation for $t-t'-V$ model show that when $t'$ is in a proper range ($\vert t'\vert \gtrsim 0.25$), for repulsive interaction, the half-filled CO phase will expand to a finite region.$\cite{JW29}$ Results from iPEPS show that CO appears only at half-filling.$\cite{PC39.1}$
Extending the calculation in this work to the $t-t'-V$ model will provide a reference for the phase diagram of this system.

Fourthly, 
the incommensurate CO is widely present in electronic materials.\cite{MA1} For the spinless fermion model, incommensurate CO was found in infinite spatial dimensions\cite{GU25} and in two-dimensional anisotropic nearest-neighbor hopping system$\cite{GM40}$ at appropriate doping. In the present study, the real space basis calculation allows all possible ways of spontaneously breaking the translational symmetry. The influence of boundary condition is weak given the large lattice size. Therefore, the fact that we did not find an incommensurate CO supports that it is unstable towards PS for fermions away from half filling, as far as the HF and p-5 basis are concerned. From the correlation point of view, however, since the p-5 basis only contains short-range correlation and partial particle-hole excitations, for the moment we cannot exclude the possibility of incommensurate ordering in the true ground state, since longer range correlation and particle-hole fluctuation could favour the incommensurate ordering. This issue deserves further study in the future.

Finally, the PS between SC and normal phases is also an interesting issue. Recently, Partridge $et$ $al$. observed the SC-normal coexistence phase in cold atom experiments with mismatched chemical potential of $^{6}Li$ atoms with different spin orientations.$\cite{GP41}$ In the SF model studied in this work, we obtain only the SC-SC (or normal-normal) PS for the attractive interaction. To study the SC-normal PS phenomenon, we need to extend our study to models of interacting fermions with spin degrees of freedom.

\end{section}

\begin{section}{Acknowledgments }

This work is supported by NSFC (Grant No.11974420).
We are grateful to helpful discussions with Q. Han and F. Yang.

\end{section}

\appendix{}

\section{ Constructing PH symmetric matrix $\bf M$ }
In PPA, Liouville matrix $\bf L$ is approximated as $\bf L \approx L(I,M)$. This approximation usually violates the PH symmetry. To restore this symmetry, we need to use a natural closed matrix $\bf M$ satisfying the PH symmetry requirement for self-consistent calculation. 

Following the idea in Ref.\onlinecite{PF14},
we use the following strategy to construct the matrix $\bf M$. 
Firstly, we divide Hamiltonian $H$ into odd and even parts under PH transformation Eq.($\ref{eq:34.1}$).
\begin{align}\label{eq:40}
&	H_e=-t\sum_{<ij>}(c_i^+c_j+h.c.)+V\sum_{<ij>}n_in_j - 2V \sum_i n_i + c, \nonumber \\ 
& H_o= (2V-\mu) \sum_i n_i - c .
\end{align}
Here, $H_e=\frac{H+H'}{2}$ and $H_o=\frac{H-H'}{2}$. $c$ = $(V-\frac{\mu}{2})N$ is a constant. 
At PH symmetry point $\mu=2V$, $H_e=H$ and $H_o=0$.
Secondly, We denote
$[\vec{A}, H_e]={\bf M}_e^T\vec{A} + \vec{B}_e$ and $[\vec{A}, H_o]={\bf M}_o^T\vec{A} + \vec{B}_o$. 
Also, the results of the composite transformation of these two equations are denoted by
$\widetilde{[\vec{A}, H_e]}=\widetilde{{\bf M}}_e^T\vec{\tilde{A}} + \vec{\tilde{B}}_e$ and $\widetilde{[\vec{A}, H_o]}=\widetilde{{\bf M}}_o^T\vec{\tilde{A}} + \vec{\tilde{B}}_o$. 
It can be proved that if $\vec{\widetilde{B}}_e = -{\bf Q}\vec{B}_e$ and $\vec{\widetilde{B}}_o = {\bf Q}\vec{B}_o$, then $\widetilde{{\bf M}}_e = -{\bf Q}^{\dagger}{\bf M}_e {\bf Q}^T$ and $\widetilde{{\bf M}}_o ={\bf Q}^{\dagger}{\bf M}_o {\bf Q}^T$. ${\bf M }= {\bf M}_e + {\bf M}_o$ meets the requirement of PH symmetry.$\cite{PF14}$
Here, the composite transformation and matrix $\bf Q$ are defined as $\tilde{O}=(O')^{\dagger}$ and $\vec{\tilde{A}} = {\bf Q}\vec{A}$, respectively.

In fact, when $\vec{A}$ is closed under PH transformation, we can remove the Hermitian conjugate operation in the definition of composite transformation. Accordingly, we need to construct $\vec{B}_e$ and $\vec{B}_o$ satisfying $\vec{\widetilde{B}}_e = {\bf Q}\vec{B}_e$ and $\vec{\widetilde{B}}_o = -{\bf Q}\vec{B}_o$ respectively. In this case, $\widetilde{{\bf M}}_e = {\bf Q}^{\dagger}{\bf M}_e {\bf Q}^T$ and $\widetilde{{\bf M}}_o =-{\bf Q}^{\dagger}{\bf M}_o {\bf Q}^T$. The ${\bf M}$ matrix that satisfies PH symmetry is ${\bf M }= {\bf M}_e + {\bf M}_o$. Actually, the two different definitions of composite transformation do not bring physical discrepancy.

\end{document}